\documentclass{aa}

\usepackage{graphicx}
\usepackage{amssymb}
\usepackage{natbib}
\usepackage{epsf}

\renewcommand{\ion}[2]{#1\,{\sc #2}}

\newcommand{\lam}{$\lambda$}

\newcommand{\ecs} {erg~cm$^{-2}$~s$^{-1}$} 
\newcommand{\kms} {km\ s$^{-1}$} 
\newcommand{\deltE}{\Delta\kern-1ptE}

\newcommand{\hst}{\emph{HST}}

\newcommand{\vtl}{$\vartriangleleft$}
\newcommand{\btl}{$\blacktriangleleft$}

\title{Fe VII lines in the spectrum of RR Telescopii}

\author{P.R.\ Young\inst{1}
\and
K.A.\ Berrington\inst{2}
\and
A.\ Lobel\inst{3}}

\institute{CCLRC Rutherford Appleton Laboratory, Chilton, Didcot,
  Oxfordshire, OX11 0QX, U.K.
\and
School of Science and Mathematics, Sheffield Hallam University,
  Sheffield, S1 1WB, U.K.
\and
Smithsonian Astrophysical Observatory, 60 Garden Street, Cambridge MA
  02138, U.S.A.}

\date{Received / Accepted}

\abstract{
Thirteen transitions within the ground $3d^2$ configuration of \ion{Fe}{vii}
are identified in ultraviolet and optical spectra of the symbiotic
star RR Telescopii obtained with the STIS instrument of
the \emph{Hubble Space Telescope}. The line fluxes are compared with
theoretical data computed with the recent atomic data of
K.A. Berrington et al., and high resolution optical spectra from
\emph{VLT}/UVES are used to identify blends. Seven branching ratios
are measured, with three 
in good agreement with theory and one affected by blending. The
\lam5277/\lam4943 branching ratio is discrepant by
$>3\sigma$, indicating errors in the atomic data for the \lam5277
line. A least-squares 
minimization scheme is used to 
simultaneously 
derive the temperature, $T$, and density, $N_{\rm e}$, of the RR Tel
nebula, and the 
interstellar extinction, $E(B-V)$, towards RR Tel from the complete set of emission
lines. The derived values are: $\log\,T/{\rm K}=4.50\pm 0.23 $, 
$\log\,N_{\rm e}/{\rm cm}^{-3}=7.25\pm 0.05$, and $E(B-V)\le
0.27$. The extinction is not well-constrained by the \ion{Fe}{vii}
lines, but is consistent with the more accurate value
$E(B-V)=0.109^{+0.052}_{-0.059}$ derived here from the
\ion{Ne}{v} \lam2974/\lam1574 ratio in the STIS spectrum. 
Large differences between the K.A. Berrington et
al.\ electron excitation data and the earlier F.P. Keenan \& P.H. Norrington
data-set are demonstrated, and the latter is shown to give worse
agreement with observations.
}

\begin{document}

\maketitle

\section{Introduction}\label{sect.intro}

RR Telescopii is a slow symbiotic nova that underwent an outburst in
1944 and whose optical light curve is still fading. The nature of the
system is not fully understood but is thought to consist of a
late-type giant and a hot white dwarf of effective temperature around
142,000~K \citep{jordan94}. The white dwarf illuminates the nebula
around the system, and gives rise to an extraordinarily
rich nebular emission line spectrum, that has long been a testing
ground for atomic data.

Comparatively little atomic data is available for \ion{Fe}{vii}, on
account of the complexity of the ion (which has an open $d$ shell),
and the fact that \ion{Fe}{vii} gives rise to only weak emission lines
from the Sun's atmosphere. However, nebular plasmas under illumination
from a hot source ($\gtrsim 100,000$~K) exhibit strong emission lines
at visible wavelengths
from within the \ion{Fe}{vii} $3d^2$ ground configuration. The
transitions offer excellent plasma diagnostic possibilities
\citep[e.g.,][]{nussb82}.

In the present work we compare theoretical line emissivities
calculated from the atomic data of \citet{berr00} for  \ion{Fe}{vii}
with the observed RR Tel fluxes measured by the STIS 
instrument on board the \emph{Hubble Space Telescope} (\hst). These
observations give almost simultaneous measurements of the optical and
UV lines, and allow the \ion{Fe}{vii} atomic model to be tested.

\section{Atomic data}

The atomic data of \citet{berr00} were used to construct a model of
the \ion{Fe}{vii} ion for inclusion in the CHIANTI database\footnote{http://www.chianti.rl.ac.uk.}
\citep{dere97, young03}. The model
consists of all 9 fine structure levels in the ground $3d^2$
configuration of the ion. 
The Maxwellian-averaged collision strengths ($\Upsilon$) for all 36 transitions
amongst the 9 levels were fit with 5 point
splines, as described in \citet{dere97}, for inclusion in CHIANTI. The
fits have an accuracy of $\le 1.4$~\% over the temperature range $4.3\le \log\,T\le 6.0$.
The radiative decay rates are from
\citet{berr00}, however three of the values given in Table~3 of
\citet{berr00} are incorrect. The 1--4, 1--7 and 4--7 decay rates
should be 0.372, 0.0150, and 0.182~s$^{-1}$.

\begin{figure*}
  \resizebox{6in}{!}{\includegraphics{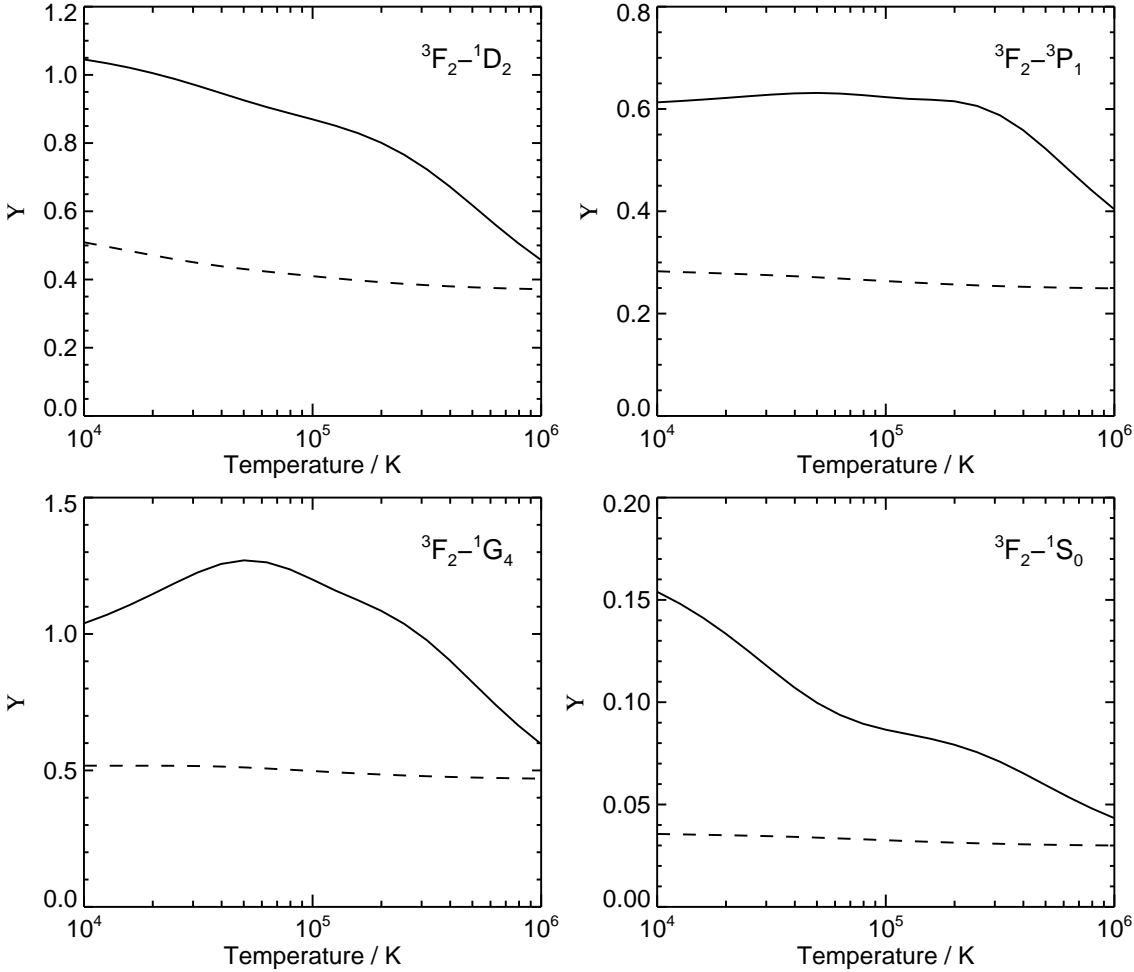}}
  \caption{Comparison of Maxwellian-averaged collision strengths
    ($\Upsilon$) from \citet{berr00} (solid line) with
    those of \citet{keenan87} (dashed line) for four transitions of
    \ion{Fe}{vii}. The values have been derived from the spline fits
    in CHIANTI. The original data cover $8000\le T\le 120,000$~K
    \citep{keenan87} and $4.3\le\log\,T\le 6.0$ \citep{berr00} and so
    values outside these ranges have been extrapolated.} 
  \label{fig.ups-compare}
\end{figure*}

The previous model of \ion{Fe}{vii} used in CHIANTI made use of
electron collision data from \citet{keenan87} and radiative decay
rates from \citet{nussb82}. Large discrepancies are found with the
data of \citet{keenan87} as illustrated in Fig.~\ref{fig.ups-compare}
for four transitions from the ground $^3F_2$ level. The differences
are due to resonance structure not accounted for by \citet{keenan87}
-- see \citet{berr00} for more details.

\begin{figure}
  \resizebox{\hsize}{!}{\includegraphics{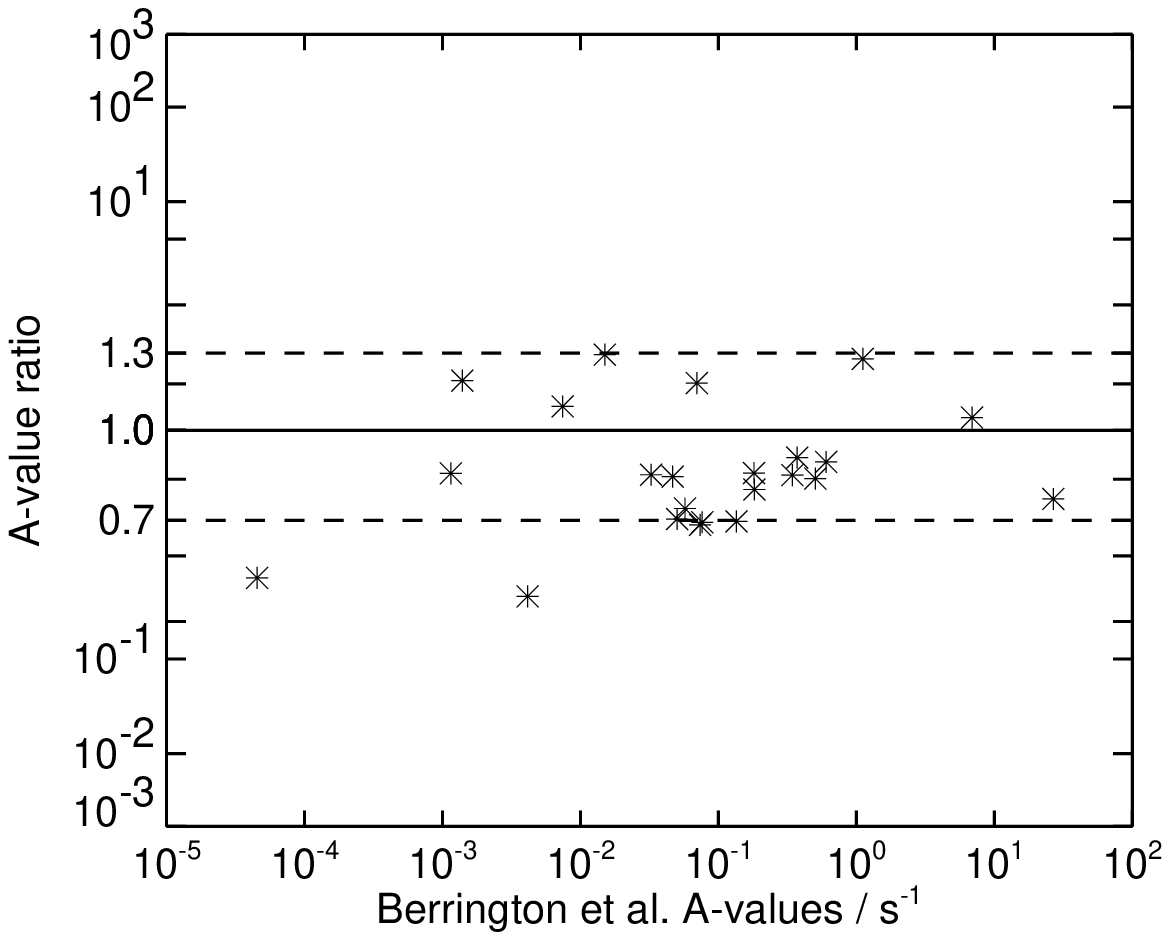}}
  \caption{Comparison of $A$-values from \citet{berr00} with those of
    \citet{nussb82}. The Y-axis gives the ratio of the \citet{nussb82}
    value to the \citet{berr00} for a particular transition, while the
    X-axis value gives the \citet{berr00} value. The Y-axis is scaled as
    the square root of the logarithm (base 10) of the ratio, allowing a
    wide range of values to be displayed. The solid line indicates
    perfect agreement (ratio=1), and the dashed lines are placed at
    $\pm$30~\%.} 
  \label{fig.A-compare}
\end{figure}

The radiative decay rates ($A$-values) of \citet{nussb82} are in very good
agreement with those of \citet{berr00}, as shown in
Fig.~\ref{fig.A-compare}. All but five of the 24 $A$-values agree to
within 30~\%. The worst agreement is for the $^1D_2$--$^1G_4$ and
$^3P_2$--$^1G_4$ transitions where the \citet{berr00} values are
factors 3.4 and 2.6 greater than those of \citet{nussb82}. Both transitions
are weak, however, and the decays to the $^3F_{3,4}$ are more
important in de-populating the  $^1G_4$ level.

\citet{trabert03} describe a laboratory measurement of the lifetime of
the 
$^1S_0$ level, giving $29.6\pm 1.8$~ms. This compares very well with
28.7~ms from the \citet{berr00} calculation. The \citet{nussb82}
calculations gave 33.1~ms.

Software available in the CHIANTI database allows level populations
and line emissivities to be calculated over a wide range of
temperature and density \citep{dere97}. In Table~\ref{tbl.lev-pops}, 
percentage level populations for the nine levels of \ion{Fe}{vii} are given
at two temperatures and three
densities. 25,000~K is a typical temperature for photoionized plasmas,
while 250,000~K is the temperature of maximum ionization of
\ion{Fe}{vii} in an electron ionized plasma.

\begin{table*}[h]
\caption{Percentage level populations for \ion{Fe}{vii} calculated at
  two values 
  of the electron temperature and three values of the electron density
  ($N_{\rm e}=10^6, 10^8, 10^{10}$~cm$^{-3}$).}
\begin{flushleft}
\begin{tabular}{llrrrrrrr}
\hline
\hline
\noalign{\smallskip}
&&\multicolumn{3}{c}{25,000~K} & &\multicolumn{3}{c}{250,000~K}\\
\cline{3-5}\cline{7-9}
\noalign{\smallskip}
Index &Level & $10^6$ &$10^8$ &$10^{10}$ &&$10^6$ &$10^8$ &$10^{10}$\\
\hline
\noalign{\smallskip}
1 & $^3F_2$ &43.05 &20.29 &19.45 &
            &62.04 &14.85 &12.06 \\
2 & $^3F_3$ &34.60 &26.79 &25.63 &
            &26.19 &21.87 &16.79 \\
3 & $^3F_4$ &19.69 &31.12 &30.60 &
            &9.16 &24.62 &21.39 \\
4 & $^1D_2$ &0.56 &6.31   &7.10 &
            &0.40 &8.70 &10.93 \\
5 & $^3P_0$ &0.37 &1.17 &1.23 &
            &0.46 &2.05 &2.14 \\
6 & $^3P_1$ &0.81 &3.43 &3.60 &
            &0.88 &6.24 &6.43 \\
7 & $^3P_2$ &0.64 &5.30 &5.71 &
            &0.52 &9.40 &10.62 \\
8 & $^1G_4$ &0.30 &5.57 &6.61 &
            &0.35 &12.15 &18.20 \\
9 & $^1S_0$ &$9.82(-5)$ &0.02 &0.08 &
            &$5.45(-4)$ &0.10 &1.44 \\
\hline
\end{tabular}
\end{flushleft}
\label{tbl.lev-pops}
\end{table*}

\section{Observations}\label{sect.obs}

RR Tel was observed on 2000 October 18 with the STIS instrument on
\hst. Complete spectra were obtained over the 
wavelength range 1140--7051\,\AA\ in 3 \hst\ orbits. The UV spectra
(1140--3120\,\AA) were obtained at a spectral resolution of around
30--45,000, while the visible spectra (3022-7051\,\AA) were obtained at a
resolution of 5--10,000.

STIS data are processed automatically  with the most recent version
of the STIS calibration pipeline when accessed through the
MAST\footnote{Multimission Archive at Space Telescope,
  http://archive.stsci.edu.} 
archive. 1D spectra are provided with fluxes and flux errors. Absolute
fluxes are accurate to 5~\% (CCD spectra) and 8~\% (MAMA spectra);
absolute wavelengths are accurate to 3--15~\kms\ (CCD spectra) and
1.5--5.0~\kms\ (MAMA spectra).

The \ion{Fe}{vii} lines are exceptionally strong in RR Tel, and 13
lines can be identified, which are listed in
Table~\ref{line-list}. The lines were fit with Gaussian profiles
through $\chi^2$ minimization using the MPFIT curve
fitting tools
of
C.B.~Markwadt\footnote{http://astrog.physics.wisc.edu/$\sim$craigm/idl/idl.html.}
as distributed through \emph{Solarsoft}\footnote{\emph{Solarsoft} is
  a set of  integrated software libraries, databases, and system
  utilities that provide a common programming and data analysis
  environment for Solar Physics. It is available at
  http://www.lmsal.com/solarsoft.}.
For 11 of the 13 lines, fitting errors were smaller than the accuracy
of the STIS absolute calibration, and so the line fluxes are accurate
to 5 and 8~\% 
for lines in the CCD and MAMA spectra, respectively. The weak \lam2183 line
has a significant fitting error, and the flux is $2.4\pm 0.4 \times
10^{-14}$~\ecs. The \ion{Fe}{vii} \lam2143 line is blended with \ion{N}{ii}
\lam2143.452. The 
nearby \ion{N}{ii} \lam2139.687 line shares a common upper level with
the latter, and the branching ratio \lam2139/\lam2143 is 0.406. The
\lam2139 flux is $7.87\times 10^{-14}$~\ecs, and by correcting for the
branching ratio and subtracting this from the blended feature's flux,
we derive an estimate of the \ion{Fe}{vii} line flux as $19.4\pm 3.3
\times 10^{-14}$~\ecs.

The low resolution optical spectra from STIS do not allow line
blending to be studied. For this purpose we have analysed high
resolution optical spectra obtained with the UVES instrument on the
\emph{Kueyen} telescope of the \emph{VLT} in 1999 October. These
spectra are not simultaneous with the STIS spectra and we note that
emission line fluxes have been found to vary in time for RR Tel
\citep{thack77,crawford99}, with low ionization lines becoming weaker
and high ionization lines becoming stronger. The 12 month separation of
the UVES and STIS spectra would have led to relative changes in
emission line fluxes of at most 10~\%, and the blending of \ion{Fe}{vii}
lines with \ion{Fe}{ii} lines discussed below will be overestimated
by at most this amount.

\begin{figure*}
  \resizebox{\hsize}{!}{\includegraphics{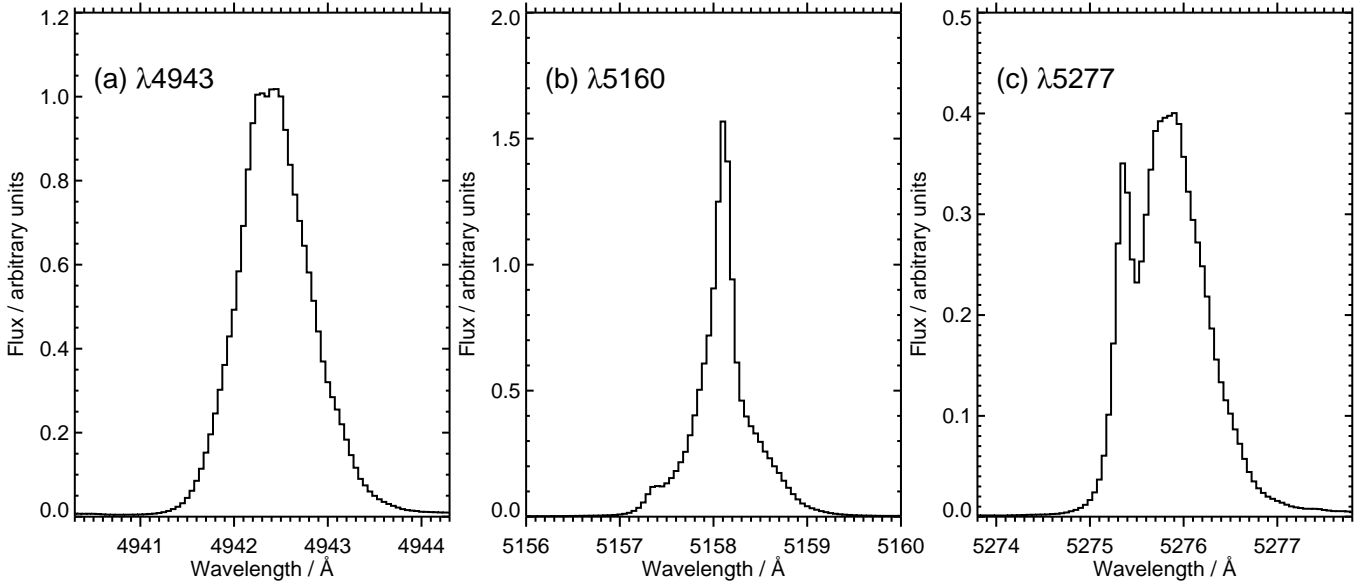}}
  \caption{Spectra from UVES showing three \ion{Fe}{vii} lines. The
    \lam4943 line is unblended, while both \lam5160 and \lam5276 are
    blended with cooler species.} 
  \label{fig.uves}
\end{figure*}

The UVES spectra
were reduced using the MIDAS package by one of us (A.~Lobel). The
\ion{Fe}{vii} optical lines are intrinscially asymmetric
(Fig.~\ref{fig.uves}a), and two of the lines clearly demonstrate
blending: \lam5160 and \lam5277 (Fig.~\ref{fig.uves}b,c). For
\lam5160, we identify the blending line as \ion{Fe}{ii} $3d^7$ $a^4F_{9/2}$ --
$3d^6(^3H)4s$ $a^4H_{13/2}$ which has a vacuum wavelength of
5160.214\,\AA. Although this is a forbidden transition (and thus has a
small $A$-value), it is the only transition from the $a^4H_{13/2}$
upper level, and thus gives rise to a strong emission line. The
\ion{Fe}{vii} \lam4894/\lam5160 
branching ratio suggests a \ion{Fe}{ii} contribution of 22~\%, and this
is consistent with the appearance of the UVES profile
(Fig.~\ref{fig.uves}b). As there are no branching ratios for the
\ion{Fe}{ii} $a^4H_{13/2}$ level then it is not possible to estimate
the \ion{Fe}{ii} contribution to \lam5160 independently of the
\ion{Fe}{vii} lines.

The \ion{Fe}{vii} \lam5277 line is blended on the short wavelength side
with an \ion{Fe}{ii} transition that we identify as 
$3d^6(^3G)4s$ $a^4G_{9/2}$ -- $3d^6(^5D)4p$ $z^4F_{7/2}$, with
vacuum wavelength 5277.480\,\AA. Another line
from this upper level has a wavelength 4557.178\,\AA\ and is
identified in the STIS spectrum with a flux $5.5\pm 0.7\times
10^{-14}$~\ecs. The branching ratio (using radiative decay rates
from the CHIANTI database) implies that the \lam5277.480
line has a flux $7.4\pm 0.9\times
10^{-14}$, thus contributing 8.7~\% to the blended STIS feature. This
is consistent with the high resolution UVES spectrum (Fig.~\ref{fig.uves}c).

\citet{selvelli01} list the \ion{Fe}{vii} \lam3587 and \lam6087 lines
as blended with \ion{Fe}{vi} and \ion{Ca}{v} lines, respectively. The
blending \ion{Fe}{vi} line (corresponding to ground configuration
transition $^4P_{1/2}$--$^2F_{5/2}$) has vacuum wavelength 3588.680~\AA, i.e.,
1.339~\AA\ from the \ion{Fe}{vii} line, and so would be easily resolvable
from the \ion{Fe}{vii} line in the high resolution UVES spectra, yet
it is not seen. Further we note that the \ion{Fe}{vi} transition shares
a common upper level with the 2145.756~\AA\ transition (decays to
$^4F_{3/2}$), and the 
branching ratio \lam3588/\lam2145 is $2.5\times 10^{-3}$ using the
radiative decay rates from CHIANTI. The \lam2145 line
is found in the STIS spectra with a flux $4\times 10^{-14}$~\ecs,
implying a \lam3588 flux of $1 \times 10^{-16}$~\ecs and thus it makes
no significant
contribution to the \ion{Fe}{vii} \lam3587 line.

The \ion{Ca}{v} line is the $^3P_1$--$^1D_2$ ground configuration
transition with vacuum wavelength 6088.058~\AA\ and is thus 0.593~\AA\
shortward of the \ion{Fe}{vii} line. The large width of the feature at
this wavelength means the two lines would be blended. Another
\ion{Ca}{v} line at 5309.580~\AA\ shares the $^1D_2$ upper level and
the theoretical ratio is 0.19 (using radiative decay rates from the
NIST database). The \lam5309 line is found in the STIS spectra with a
flux $7.94\times 10^{-13}$~\ecs, thus implying  a \lam6088 flux of
$1.51\times 10^{-13}$~\ecs\ -- only a 1.7 \% contribution to the
\lam6087 feature in the STIS spectrum. Correcting for this leads to a
flux for the 
\ion{Fe}{vii} component of $8.56 \times 10^{-12}$~\ecs.

The high resolution of the STIS/MAMA spectra allow line widths to be
measured, and the \lam2016 line has a width of 53~\kms\ (0.357~\AA). A
correlation between line width and ionization potential has been noted
before in RR Tel \citep{penston83}. The ionization potential of
Fe$^{+6}$ is 99.1~eV, and the \lam2016 line width
compares well with the widths of lines of Ne$^{+4}$ (97.1~eV) and
Mg$^{+4}$ (109.3~eV): the \ion{Ne}{v} \lam1574 line has a width of
46~\kms, and the \ion{Mg}{v} \lam1324 line has a width of 50~\kms,
both measured from the STIS spectra.

\begin{figure*}
  \resizebox{\hsize}{!}{\includegraphics{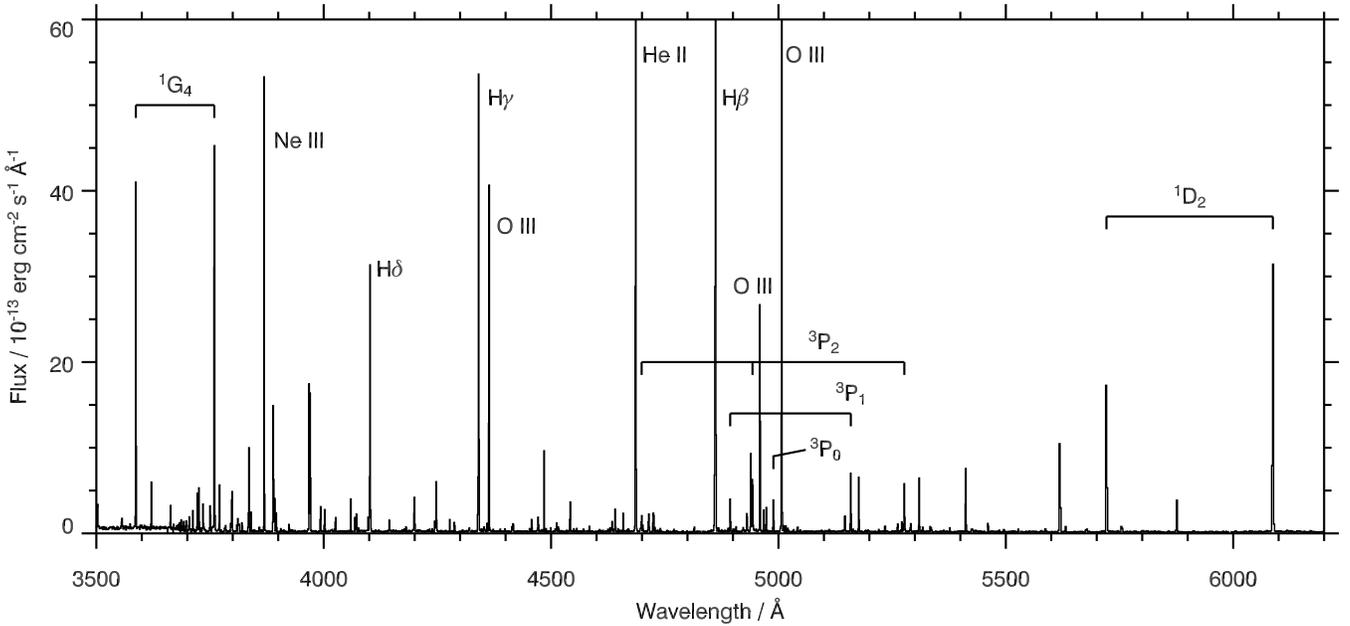}}
  \caption{A section of the optical spectrum of RR Tel from \hst/STIS,
    covering 3500--6200~\,\AA. The ten \ion{Fe}{vii} lines are identified
    by their upper emitting level }
  \label{fig.optical}
\end{figure*}

Fig.~\ref{fig.optical} shows part of the
optical spectrum, with the \ion{Fe}{vii} and several other prominent
lines identified. 

\begin{table}[h]
\caption{Fe\,VII line list for RR Telescopii.}
\begin{flushleft}
\begin{tabular}{lrll}
\hline
\hline
\noalign{\smallskip}
$\lambda_{\rm meas}$ (\AA) &Flux$^a$ &$\lambda_{\rm vac}$ (\AA) & Transition\\
\hline
\noalign{\smallskip}
2015.570 &67.4
     &2016.015 &$^1D_2$ -- $^1S_0$ \\
2143.085$^b$ &38.8
     &2143.706 &$^3P_1$ -- $^1S_0$ \\
2182.902 &2.4
     &2183.421 &$^3P_2$ -- $^1S_0$ \\
3586.830 &555.6
     &3587.341 &$^3F_3$ -- $^1G_4$ \\
3759.526 &838.0
     &3759.992 &$^3F_4$ -- $^1G_4$ \\
4699.118 &31.5
     &4699.557& $^3F_2$ --  $^3P_2$ \\
4893.933 &58.5
     &4894.739& $^3F_2$ --  $^3P_1$ \\
4942.957 &127.4
     &4943.863 & $^3F_3$ -- $^3P_2$ \\
4988.927 &59.9
     &4989.954 &$^3F_2$ -- $^3P_0$ \\
5159.088$^c$ &105.7
     &5160.332& $^3F_3$ --  $^3P_1$ \\
5276.792$^d$ &85.0
     &5277.853 &$^3F_4$ -- $^3P_2$ \\
5721.406 &533.7
     &5722.297& $^3F_2$ --  $^1D_2$  \\
6087.220$^e$ &870.9
     &6088.651& $^3F_3$ --  $^1D_2$  \\
\hline
\noalign{\smallskip}
\multicolumn{4}{l}{$^a$ units: $10^{-14}$ \ecs.}\\
\multicolumn{4}{l}{$^b$ blended with \ion{N}{ii} \lam2143.452.}\\
\multicolumn{4}{l}{$^c$ blended with \ion{Fe}{ii} \lam5160.214.}\\
\multicolumn{4}{l}{$^d$ blended with \ion{Fe}{ii} \lam5277.480.}\\
\multicolumn{4}{l}{$^e$ blended with \ion{Ca}{v} \lam6088.058.}\\
\end{tabular}
\end{flushleft}
\label{line-list}
\end{table}

\section{Comparison with observations}

\subsection{Branching ratios}\label{sect.branch}

For emission lines that share a common upper level in the ion, the
line ratio is 
insensitive to the plasma conditions in an optically thin plasma and
is proportional to the ratio of radiative decay rates for the
transitions. Such ratios thus provide a means for checking the quality
of radiative data for an ion. Seven such ratios are found in the STIS
spectra of RR Tel, and are given in Table~\ref{branch}. Branching
ratios in astrophysical spectra are affected by interstellar
extinction -- wavelength dependent absorption and scattering by
interstellar dust grains -- usually quantified by the quantity
$E(B-V)$. A value of 0 corresponds to no extinction, while 0.10 is the
value typically assumed for RR Tel
\citep[e.g.,][]{penston83}. Table~\ref{branch} thus gives the
theoretical ratio values calculated for these two extinction values,
using the extinction curve parameterisation of \citet{fitz99}.

Only three of the seven ratios agree with the theoretical values within the
1$\sigma$ error bars on the fluxes. Of the discrepant ratios,
\lam4894/\lam5160 is affected by the blend to the \lam5160 line
mentioned in the previous section which accounts for the observed
ratio exceeding the theoretical value. The discrepancy for the
\lam5277/\lam4943 ratio can not be resolved by blending in the \lam4943
line as this would
then make the \lam4699/\lam4943 disagree with theory. Future
calculations of radiative decay rates should aim to resolve the
problems highlighted here.

\begin{table*}[h]
\caption{\ion{Fe}{vii} branching ratios with observed values from the RR
  Tel spectrum. Two theoretical values are given: one for no
  interstellar extinction; the other for an extinction
  $E(B-V)=0.10$. \vtl\ denotes that the ratio is discrepant with theory by
  $> 
  1\sigma$; \btl\ a discrepancy of $> 3\sigma$.}
\begin{flushleft}
\begin{tabular}{lllll}
\hline
\hline
\noalign{\smallskip}
Upper &&&\multicolumn{2}{c}{$E(B-V)$} \\
\cline{4-5}
\noalign{\smallskip}
level &Ratio & Observed &0.00 &0.10 \\
\hline
\noalign{\smallskip}
$^1S_0$ &\lam2143$^a$/\lam2016 &$0.288\pm 0.054$
        &0.242 &0.224  \vtl \\
        &\lam2183/\lam2016 &$0.036\pm 0.007$ 
        &0.038 &0.036 \\
$^1G_4$ &\lam3587/\lam3759 &$0.663\pm 0.047$ 
        &0.715 &0.704 \\
$^3P_2$ &\lam4699/\lam4943 &$0.247\pm 0.017$ 
        &0.226 &0.221  \vtl\\
        &\lam5277$^b$/\lam4943 &$0.609\pm 0.045$ 
        &0.988 &1.017 \btl \\
$^3P_1$ &\lam4894/\lam5160$^c$ &$0.553\pm 0.039$
        &0.695 &0.678  \vtl\\
$^1D_2$ &\lam5722/\lam6088$^d$ &$0.623\pm 0.044$ 
        &0.656 &0.642 \\
\hline
\multicolumn{5}{l}{$^a$ flux corrected for blend with \ion{N}{ii}
  \lam2143.452.}\\ 
\multicolumn{5}{l}{$^b$ flux corrected for blend with \ion{Fe}{ii} \lam5277.480.}\\
\multicolumn{5}{l}{$^c$ blended with \ion{Fe}{ii} \lam5160.214 (not corrected).}\\
\multicolumn{5}{l}{$^d$ flux corrected for blend with \ion{Ca}{v} \lam6088.058.}\\
\end{tabular}
\end{flushleft}
\label{branch}
\end{table*}

\subsection{Derivation of physical parameters}

For an optically thin plasma the line ratios amongst the various
emission lines are determined entirely by the plasma temperature and
density, and 
interstellar extinction. From the STIS fluxes we can thus determine
each of these parameters. Rather than consider only individual line
ratios as is often done, we instead employ a minimization
procedure to \emph{all} of 
the observed emission lines simultaneously. This provides a much more
stringent test on the atomic data, but gives greater confidence in the
derived parameters.



The method applied is as follows: lines from the same upper level are
summed, thus giving six 
independent measured quantities corresponding to the six upper emitting
levels (see Table~\ref{line-list}). The \lam2143, \lam5277 and
\lam6087 line
fluxes have been 
corrected for line blending following the discussion in
Sect.~\ref{sect.obs}, but  the \lam5160 
is not included in the analysis as the blending contribution can not
be estimated 
independently of the \lam4894/\lam5160 branching ratio.
The summed line fluxes for each level are then divided through by the
summed flux for the $^1D_2$ level, leading to five independent
measurables.

A least-squares minimization procedure was applied using the MPFIT
package such that the 
temperature is allowed to  vary over the range $3.9\le \log\,T/{\rm K}\le
5.5$, the electron density over $5.0\le \log\,N_{\rm e}/{\rm cm}^{-3}\le
9.0$, and the extinction  $0.0\le E(B-V)\le 1.0$. At
each iteration, the level balance equations for \ion{Fe}{vii} are solved
at the temperature and density values using the CHIANTI software, and emissivities for each line
calculated. These are then adjusted for the extinction value, and line
ratios compared with the observed values.

The final results of the minimization procedure were $\log\,T/{\rm
  K}=4.50\pm 0.23$, $\log\,N_{\rm e}/{\rm cm}^{-3}=7.25\pm 0.05$ and
$E(B-V)=0.11\pm 
0.16$. The $\chi^2$ value was 2.3. 
A comparison between the observed
ratios and the fitted values is given in
Table~\ref{tbl.fit}. Agreement is good, with three ratios lying
within the error bars on the data.
The discrepancy for the $^3P_2$ level is likely
due to the branching ratio problem for the $^3P_2$ level noted in
Sect.~\ref{sect.branch}. 

The $E(B-V)$ value is poorly
constrained due to the fact that low values of the extinction cause
relatively little change in line fluxes even over the wide wavelength
coverage of the \ion{Fe}{vii} lines. Performing the minimization with
$E(B-V)$ fixed at 0.10 gives temperature and density values of
$\log\,T/{\rm K}=4.50\pm 0.02$, $\log\,N_{\rm e}/{\rm cm}^{-3}=7.25\pm
0.05$.

\begin{table}[h]
\caption{Comparison of observed line ratios with those derived from
  the fitting procedure. The ratios are formed by summing lines from a
  common upper level, and comparing with the summed lines from the
  $^1D_2$ 
  level.}
\begin{flushleft}
\begin{tabular}{lll}
\hline
\hline
\noalign{\smallskip}
Upper &\multicolumn{2}{c}{Ratio}\\
\cline{2-3}
\noalign{\smallskip}
level &Observed &Fitted \\
\hline
\noalign{\smallskip}
  $^3P_0$ &  $0.043 \pm   0.003$ &  0.039 \\
  $^3P_1$ &  $0.042 \pm   0.003$ &  0.042 \\
  $^3P_2$ &  $0.170 \pm   0.008$ &  0.183 \\
  $^1G_4$ &  $1.003 \pm   0.051$ &  1.016 \\
  $^1S_0$ &  $0.064 \pm   0.005$ &  0.063 \\
\hline
\end{tabular}
\end{flushleft}
\label{tbl.fit}
\end{table}

This same minimization procedure was repeated using the previous
CHIANTI dataset, but the solution did not converge within the prescribed
parameter bounds, demonstrating that the \citet{keenan87} data does
not accurately reproduce the observed fluxes.

Previous estimates of the RR Tel nebula temperature through emission
line diagnostics give values of 10--25,000~K \citep{penston83,
hayes86, doschek93, espey96}, consistent with the \ion{Fe}{vii} result
within the error bars.

Densities previously derived for the RR Tel nebula vary widely with 
ionization species, as best illustrated by Table~6 of \citet{hayes86},
where values $4.9\le \log\,N_{\rm e}/{\rm cm}^{-3} \le 9.0$ are given. All the
species studied there, however, are low ionization and so not comparable to
\ion{Fe}{vii}. \citet{espey96} adopt a density of $\log\,N_{\rm e}/{\rm cm}^{-3}=6.0$
from an analysis of  \ion{Ne}{v} and \ion{Ne}{vi} lines in spectra 
from the
\emph{Hopkins Ultraviolet Telescope} (\emph{HUT}). The ionization potentials of
these ions are 97.1 and 126.2~eV, compared to 99.1~eV for
Fe$^{+6}$, and so the ions are comparable. We
note that the \emph{HUT} spectra are only of moderate spectral resolution, and
high resolution spectra from \emph{FUSE} would be valuable in
improving the density estimate from the \ion{Ne}{vi} lines.

Previous estimates for the extinction towards RR Tel include the value
of $E(B-V)=0.10$ derived from \ion{He}{ii} recombination lines in IUE
spectra \citep{penston83}, and $0.08\pm 0.03$ from the \ion{Ne}{v}
\lam2974/\lam1574 branching ratio \citep{espey96}. This latter pair of
lines are found in the STIS spectrum, and we find an observed ratio of
$0.469\pm 0.053$. The theoretical ratio is 0.376 \citep{galavis97},
leading to an 
extinction of $0.109^{+0.052}_{-0.059}$, using the extinction curve
parameterisation of 
\citet{fitz99}. This is consistent with the upper limit (0.27) found
here from the \ion{Fe}{vii} lines.

\section{Conclusions}

The $3d^2$ ground configuration transitions of \ion{Fe}{vii} provide
excellent plasma diagnostic opportunities for nebular plasmas
illuminated by hot stars. The most recent \ion{Fe}{vii} electron
excitation data from \citet{berr00} are a great improvement over the 
previous data-set of \citet{keenan87} and the complete set of
\ion{Fe}{vii} lines in the \hst/STIS spectra of RR Tel yield 
values for the electron temperature and density, and extinction.
The temperature and extinction are consistent with values derived from
other ions, while we believe the density estimate is the most accurate
obtained so far for RR Tel at ionization levels of around 100~eV.
Four of the seven \ion{Fe}{vii} branching ratios do not agree with the
observed values within the error bars. One of these ratios
(\lam4894/\lam5160) is affected by blending in the \lam5160 line and
the UVES profile for the \lam5160 is consistent with the blending
\ion{Fe}{ii} line accounting for the blend.


The
accurate flux calibration of STIS, coupled with the strong
\ion{Fe}{vii} lines found in RR Tel, make the STIS RR Tel spectrum
discussed here an excellent benchmark for checking the accuracy of
atomic physics calculations for \ion{Fe}{vii}. 

The atomic data used here will be included in version~5 of the CHIANTI
atomic database.

\begin{acknowledgements}

A.L. acknowledges financial support provided by
STScI grants HST-GO-09369.01-A and HST-GO-10212.01-A, 
and by NASA FUSE grants GI-D107 and GI-E068.

\end{acknowledgements}

\end{document}